\documentclass[12pt]{article}
\usepackage[a4paper]{geometry}
\usepackage{amsfonts,amsmath,amssymb}
\usepackage{mathrsfs}
\usepackage{color}
\usepackage{theorem,cite}

\newtheorem{thm}{Theorem}[section]
\newtheorem{prop}[thm]{Proposition}

\newtheorem{cor}[thm]{Corollary}

\theorembodyfont{\rmfamily}
\newtheorem{rmk}{Remark}[section]

\newcommand{\bea}{\begin{eqnarray}}
\newcommand{\eea}{\end{eqnarray}}
\newcommand{\beali}{\begin{align}}
\newcommand{\eeali}{\end{align}}
\newcommand{\beano}{\begin{eqnarray*}}
\newcommand{\eeano}{\end{eqnarray*}}
\newcommand{\beq}{\begin{equation}}
\newcommand{\eeq}{\end{equation}}

\newcommand{\mb}[1]{\hspace{2.1ex}\mbox{#1}\hspace{2.1ex}}

\def\ft{{\mathfrak t}}

    \def\cF{{\cal F}}
    
    \def\cL{{\cal L}}

    \def\cU{{\cal U}}
     
\def\cY{{\cal Y}}  

\newcommand{\CC}{{\mathbb C}}

\newcommand{\HH}{{\mathbb H}}
\newcommand{\II}{{\mathbb I}}

\newcommand{\ZZ}{{\mathbb Z}}

\newcommand{\wh}[1]{\widehat{#1}}
\newcommand{\wt}[1]{\widetilde{#1}}

\newcommand{\half}{{\textstyle{\frac{1}{2}}}}
\newcommand{\sfrac}[2]{{\textstyle{\frac{#1}{#2}}}}
\newcommand{\gelpn}{{\mathcal{A}_{q,p}(\widehat{gl}(N)_{c})}}
\newcommand{\gelp}{{\mathcal{A}_{q,p}(\widehat{gl}(2)_{c})}}
\newcommand{\elp}{{\mathcal{A}_{q,p}(\widehat{sl}(2)_{c})}}
\newcommand{\car}[2]{\begin{bmatrix}{#1}\\{#2}\end{bmatrix}}
\newcommand{\proof}{\textbf{Proof:}~}
\newcommand{\qed}{{\hfill \rule{5pt}{5pt}}\\}

\def\qdet{\mbox{\rm q-det}\,}
\def\eps{\varepsilon}

\def\tr{\mathop{\rm tr}\nolimits}
\def\sgn{\mathop{\rm sgn}\nolimits}

\newcommand{\ellipt}[1]{\mbox{\AA${}_{q,p,c}(\wh{gl}_{#1})$}}

\numberwithin{equation}{section}

\begin{document}
\pagestyle{empty}

\null
\vspace{20pt}

\parskip=6pt

\begin{center}
\begin{LARGE}
\textbf{Deformed Virasoro algebras}

\textbf{from elliptic quantum algebras}
\end{LARGE}

\vspace{50pt}

\begin{large}
{J.~Avan${}^{a}$, L.~Frappat${}^b$, E.~Ragoucy${}^b$ \footnote[1]{avan@u-cergy.fr, luc.frappat@lapth.cnrs.fr, eric.ragoucy@lapth.cnrs.fr}}
\end{large}

\vspace{15mm}

${}^a$ \textit{Laboratoire de Physique Th\'eorique et Mod\'elisation (CNRS UMR 8089),} \\
\textit{Universit\'e de Cergy-Pontoise, F-95302 Cergy-Pontoise, France} \\

\vspace{5mm}

${}^b$ \textit{Laboratoire de Physique Th\'eorique LAPTh, CNRS -- Universit\'e Savoie Mont Blanc,} \\
\textit{BP 110, F-74941 Annecy-le-Vieux Cedex, France}

\end{center}

\vspace{4mm}

\begin{abstract}
We revisit the construction of deformed Virasoro algebras from elliptic quantum algebras of vertex type, generalizing the  bilinear trace procedure proposed in the 90's. It allows us to make contact with the vertex operator techniques that were introduced separately at the same period. As a by-product, the method pinpoints two critical values of the central charge for which the center of the algebra is extended, as well as (in the $gl(2)$ case) a Liouville formula.
\end{abstract}

\vfill
\rightline{LAPTH-034/16\qquad\qquad}
\vfill

\newpage

\baselineskip=16.5pt
\parskip=6pt
\parindent=0pt
\pagestyle{plain}
\setcounter{page}{1}

\section{Introduction}

Several years ago the construction and subsequent investigation of Deformed Virasoro Algebras (DVA) and their generalization to higher-rank as Deformed $W_N$ Algebras (DWA), were the object of a wealth of studies, particularly in connection with various aspects of the theory of quantum groups and quantum algebras. 
Their relevance in several algebraic-based problems in mathematical physics
was pointed out throughout these last 20 years: links with Macdonald polynomials extending to Ruijsenaars--Schneider integrable $N$-body dynamics \cite{AKOS,SKAO}; symmetries of restricted SOS models \cite{LP}; construction of ZF algebras for large $N$ limit of XYZ spin chains \cite{Luk96}; extensions of the AGT duality \cite{AGT} to 5-dimensional $N=2$ superconformal gauge theories \cite{AwaYam}.

The notion of DVA and DWA covers in fact a number of non-linear algebraic structures, characterized by quadratic exchange relations and the existence of one or several continuous free parameters. 
We restrict ourselves to the two-parameter $(p,q)$ deformation \cite{SKAO} canonically known as DVA. 
Multi-continuous parameter deformations of Virasoro algebra known diversely as Elliptic Virasoro algebra \cite{Nieri} and $A_2^{(2)}$ DVA \cite{BraLu} have appeared in the literature. 
They may have some connection with elliptic quantum algebras but we shall not address this issue here.
Scaling separately $p$ and $q$ to 1 allows in the DVA case to recover the non-deformed structure of Virasoro algebra. 
In addition there exists a notion of semiclassical limit whereby the exchange relation between a priori quantum 
operators becomes trivial when another independent parameter (usually denoted $\hbar$) is sent to zero. 
Expansion of the exchange relation around this zero-value allows one to define a classical Poisson structure, which should be consistently interpreted as a $q$-deformation of the linear classical Virasoro or $W_n$ Poisson algebra provided that $q$ and $\hbar$ be independent.

Originally discussed in e.g. \cite{CZ} the quantum DVA was indeed built as a quantization of a classical structure defined on the extended center of the affine quantum algebra $U_q (sl(2))_c$ at $c=-2$, $q$ being the deformation parameter \cite{RSTS}. 
This construction was extended to classical $q$-$W$ algebras in \cite{FR}.
The procedures to explicitly build realizations of DVA and DWA are multifold (see e.g. \cite{Das} for an example) but we will concentrate here on two types most relevant for our purpose. 
One is based on an extension of the current algebra construction of non-deformed Virasoro and $W_N$ algebras by $q$-deformation of the original Miura transformation \cite{FF}. 
In the approach \cite{FF,AKOS} the construction is explicitly shown to generate the DVA from an ($R$-matrix parametrized) ZF algebra. 
It exhibits also an intriguing connection to MacDonald polynomials \cite{McD}, hence indirectly to quantum Ruijsenaars--Schneider models of which they are eigenfunctions \cite{Rui}. 

The second one was developed by several authors in a series of papers \cite{AFRS97b,AFRS99,AFRS97a}. 
One starts from the Lax operator (R-matrix) realization of the elliptic quantum algebra $\gelpn$ \cite{FIJKMY,Fron,JKOS,ABRR} and parallel at this more abstract level the construction in \cite{RSTS}: one defines generating operator functionals of the algebra DVA/DWA, as auxiliary space traces of bilinear forms in the Lax matrices themselves. 
It is then possible to obtain at one shot both classical and quantum DVA/DWA structures, following a two-fold procedure of reduction in the space of parameters of the evaluated elliptic algebras $\{ p,q,c \}$. A first analytic relation between $p,q,c$ ensures the existence of a quadratic exchange subalgebra. A second analytic relation ensures the abelianity of this quadratic exchange subalgebra.

The constructions in \cite{AFRS97b,AFRS99} however relied on a specific form of the bilinear generating function, and hence certainly defined a non-exhaustive set of analytic conditions. 
It was particularly noticeable that the explicit quantum DVA constructed by VOA techniques in \cite{SKAO,Shi} could not be reached by the bilinear forms on Lax operator used in \cite{AFRS97b,AFRS99,AFRS97a}. 
This provided us with a strong incentive to reexamine and extend the study of the connection between $\gelpn$ and DVA/DWA.

The purpose of this work is therefore to propose a more systematic and extended formulation of the Lax matrix construction of deformed Virasoro (hopefully subsequently extensible to $W_N$) algebras within the general context of the twofold procedure, in order to get again at the same time classical and quantized algebra structures. 

The paper runs as follows. Section \ref{sect:elliptic} prepares the ground by reminding the reader of essential definitions and key properties of the elliptic algebras based on $\wh{gl}(N)_c$. 
Section \ref{sect:main} contains the bulk of results, starting from these specific elliptic algebras. 
It presents the construction of two quadratic exchange subalgebras both characterized by general analytic relations between parameters $p,q,c$, namely $(-p^{\frac{1}{2}})^m (-p^{*\frac{1}{2}})^n = q^{-N}$ for any integer $m,n \in \ZZ$. 
We identify in one specific case the  quantum DVA originally proposed in \cite{SKAO}.
We derive in Section \ref{sect:poisson} the Poisson structures as limits of DVA, recovering  the original classical DVA. 
 Section \ref{sect:proof} contains the proofs of the results presented in section \ref{sect:main}.  
Some open issues are dealt with in Section \ref{sect:conclu}.
Appendices are devoted to known results on elliptic algebras.

\section{Elliptic quantum algebras based on $\wh{gl}(N)_c$\label{sect:elliptic}}
We will deal with two versions of elliptic algebras that we present here. Both are of vertex type: the face type elliptic algebras will not be treated in this paper.

\subsection{The elliptic quantum algebra $\gelpn$\label{sect:elpn}}
We remind here the definition of the elliptic quantum algebra $\gelpn$ \cite{FIJKMY,JKOS,AFRS99}.
We consider the free associative algebra generated by the operators $L_{ij}[n]$ where $i,j \in \ZZ_N$ and $n \in \ZZ$, and we define the formal series 
\begin{equation}
L_{ij}(z) = \sum_{n\in\ZZ} L_{ij}[n] \, z^n
\end{equation}
encapsulated into a $N \times N$ matrix
\begin{equation}
L(z) = \begin{pmatrix} 
L_{11}(z) & \cdots & L_{1N}(z) \\
\vdots && \vdots \\ 
L_{N1}(z) & \cdots & L_{NN}(z) \\ 
\end{pmatrix} \,.
\end{equation}
One defines $\gelpn$ by imposing the following constraints on $L_{ij}(z)$:
\begin{equation}
\widehat R_{12}(z/w) \, L_1(z) \, L_2(w) = L_2(w) \, L_1(z) \, \widehat R_{12}^{*}(z/w) \,,
\label{eq232}
\end{equation}
where $L_1(z) \equiv L(z) \otimes \II$, $L_2(z) \equiv \II \otimes L(z)$, $R_{12}(z) \equiv R_{12}(z,q,p)$ is the $N$-elliptic $R$-matrix defined in Appendix \ref{app:B} and $\widehat R^{*}_{12}(z) = \widehat R_{12}(z,q,p^*=pq^{-2c})$. 

It is useful to introduce the following two matrices:
\begin{align}
& L^+(z) = L(q^{\frac{c}{2}}z) \,, \label{eq:defLp} \\
& L^-(z) = (g^{\frac{1}{2}} h g^{\frac{1}{2}}) \, L(-p^{\frac{1}{2}}z) \, (g^{\frac{1}{2}} h g^{\frac{1}{2}})^{-1} \,, \label{eq:defLm} 
\end{align}
where the matrices $g$ and $h$ are defined in \eqref{def:gij} and \eqref{def:h}.
They obey coupled exchange relations following from \eqref{eq232}, periodicity condition \eqref{eq225} and unitarity property \eqref{eq:unitarity} of the matrices
$\widehat R_{12}$ and $\widehat R^{*}_{12}$:
\begin{align}
\widehat R_{12}(z/w) \, L^\pm_1(z) \, L^\pm_2(w) & = L^\pm_2(w) \, L^\pm_1(z) \, \widehat R^{*}_{12}(z/w) \,, 
\label{eq234}\\
\widehat R_{12}(q^{\frac{c}{2}}z/w) \, L^+_1(z) \, L^-_2(w) & = L^-_2(w) \, L^+_1(z) \, \widehat R^{*}_{12}(q^{-\frac{c}{2}}z/w) \,.
\label{eq236}
\end{align}

\subsection{The elliptic quantum algebra \ellipt{N}\label{sect:vertex}}
Using the notations introduced in section \ref{sect:elpn}, one defines an alternative elliptic quantum algebra through the unitary $R$-matrix \eqref{eq220}, instead of the matrix $\widehat R$, i.e.
\begin{align}
R_{12}(z/w) \, \cL^\pm_1(z) \, \cL^\pm_2(w) &= \cL^\pm_2(w) \, \cL^\pm_1(z) \, R^{*}_{12}(z/w) \,, 
\label{eq:elpalt1}\\
R_{12}(q^{\frac{c}{2}}z/w) \, \cL^+_1(z) \, \cL^-_2(w) &= \cL^-_2(w) \, \cL^+_1(z) \, R^{*}_{12}(q^{-\frac{c}{2}}z/w) \,.
\label{eq:elpalt2}
\end{align}
Let us stress that although the two $R$-matrices $R$ and $\wh R$ differ only by a normalization, the associated algebras are different because of the asymmetry in $c$ in eq. \eqref{eq:elpalt2}.

Since the matrix $\wh R_{12}(z)$ is the evaluation of the universal $R$-matrix, one is led naturally to the $\gelpn$ algebra, as defined by \eqref{eq234}-\eqref{eq236}. However, the $R$-matrix \eqref{eq220} is used in the construction of free field realizations for elliptic algebras. 
Indeed, the bosonization used in \cite{AKOS9612233} is based on vertex operators that obey a Zamolodchikov-Faddeev algebra defined with the $R$-matrix \eqref{eq220}. 

We will show below that the elliptic algebra defined by relations \eqref{eq:elpalt1}-\eqref{eq:elpalt2} allows 
us to make contact between the general approach defined in the present paper and in \cite{AFRS97a,AFRS97b} with the one used in \cite{AKOS9612233}, filling the gap between these two constructions of deformed Virasoro algebras.

\section{Main results\label{sect:main}}
We present here the main results which we have achieved. To ease the reading we postpone the somewhat long and technical proofs to section \ref{sect:proof}.
 We first single out the $gl(2)$ case, where supplementary results can be obtained, and subsequently present the relevant generalizations  to the $gl(N)$ case.

\subsection{Exchange algebras in elliptic quantum algebras based on $gl(2)$\label{sect:exchange2}}
\subsubsection{Case of $\gelp$ elliptic quantum algebra\label{sect:t-n=2}}
\begin{thm}[Quadratic subalgebras in $\gelp$]\label{thmone}\ \\
In the three-dimensional parameter space spanned by $q,p,c$, we define the following two-dimensional surfaces 
\beq
\mathscr{S}_{mn}\ :\ (-p^{\frac{1}{2}})^{m} (-p^{*\frac{1}{2}})^{n} = q^{-2}\,,\quad \forall m,n \in \ZZ,\quad n \ne 0.
\eeq
We introduce the generators 
\begin{eqnarray}
t_{mn}(z) &=& \tr \Big( (g^{\frac{1}{2}} h g^{\frac{1}{2}})^{-m}\, L\big((-p^{*\frac{1}{2}})^{n}z\big)\,
 (g^{\frac{1}{2}} h g^{\frac{1}{2}})^{-n}\, L(z)^{-1} \Big) ,
\label{eq:deftmn}\\
t^*_{-n,-m}(z) &=& \tr \Big( (g^{\frac{1}{2}} h g^{\frac{1}{2}})^{n}\, L\big((-p^{\frac{1}{2}})^{-m}z\big)^{-1}\,
 (g^{\frac{1}{2}} h g^{\frac{1}{2}})^{m}\, L(z) \Big) .
\label{eq:deftmnst}
\end{eqnarray}
We remind that the matrices $g$ and $h$ are defined in \eqref{def:gij} and \eqref{def:h}.

On the surface $\mathscr{S}_{mn}$ we have the following exchange relations:
\begin{align}
&t_{mn}(z) \, L(w) \ =\  \frac{\mathcal{F}_{-m}(z/w) }{ \mathcal{F}^*_n(z/w)} \; L(w) \, t_{mn}(z),
\label{eq:exchtL}\\[1ex]
&t^*_{-n,-m}(z) \, L(w) \ =\  \frac{\mathcal{F}^*_n(z/w)}{\mathcal{F}_{-m}(z/w) } \; L(w) \, t^*_{-n,-m}(z),
\label{eq:exchtLalt}\\[1ex]
\intertext{implying the quadratic exchange relations:}
&t_{mn}(z) \, t_{mn}(w) \ =\  \mathcal{Y}_{mn}(z/w) \, t_{mn}(w) \, t_{mn}(z),
\label{eq:exchtt}\\
&t_{-n,-m}^*(z) \, t_{-n,-m}^*(w) \ =\  \mathcal{Y}_{mn}(z/w) \, t_{-n,-m}^*(w) \, t_{-n,-m}^*(z),
\label{eq:ttstar}\\
&t_{mn}(z) \, t_{-n,-m}^*(w) \ =\  \mathcal{Y}_{mn}(z/w)^{-1} \, t_{-n,-m}^*(w) \, t_{mn}(z),
\label{eq:exchttstar}
\end{align}
where 
\begin{equation}
\mathcal{F}_{a}(x) = \left\{ 
\begin{array}{ll}
\displaystyle \prod_{k=0}^{a-1} \mathcal{U}\big((-p^{\frac{1}{2}})^k x\big) & \text{for $a > 0$} 
\\[3ex]
1 & \text{for $a = 0$} \\
\displaystyle \prod_{k=1}^{|a|} \mathcal{U}\big((-p^{\frac{1}{2}})^{-k} x\big)^{-1} & \text{for $a < 0$}
\end{array}
\right. 
\qquad\mb{;} \mathcal{F}^*_a(x) = \mathcal{F}_a(x)\big\vert_{p \to p^*}.
\label{eq:exprFn}
\end{equation}
$\cU(x)$ is defined in relation \eqref{def:U}.
The function $\cY_{mn}$ can be written as
\beq\label{Y:gg}
\mathcal{Y}_{mn}(x) = \frac{g_{mn}(x^2)}{g_{mn}(x^{-2})}
\mb{with}
g_{mn}(z) = g_{corr}(z)\,\Big(\prod_{k=1}^{|m|-1} g^{(k)}(z)\Big)^2\Big(\prod_{k=1}^{|n|-1} g^{*(k)}(z)\Big)^{-2}
\eeq
where
\beq\label{def:g}
\begin{aligned}
g^{(k)}(z) &= \exp\Big(\sum_{\ell=1}^\infty \frac{(1-p^{-k\ell})(1-(p^{k}q^2)^\ell)}{1+q^{2\ell}}\, \frac{z^\ell}{\ell}\Big)
\\
g^{*(k)}(z) &= g^{(k)}(z)\Big|_{p \to p^*}
\\
g_{corr}(z) &= g^{(|m|)}(z)\,\big(g^{*(|n|)}(z)\big)^{-1}
\end{aligned}
\eeq
\end{thm}
The proof of this theorem is given in section \ref{subsect:C1}.

\begin{rmk}
On the surface $\mathscr{S}_{mn}$, the generator $t_{mn}(z)$ also exhibits quadratic exchange relations with all $t_{rs}(w)$, albeit with different structure functions: 
\begin{equation}
t_{mn}(z) \, t_{rs}(w) \ =\  \frac{\mathcal{F}^*_{n}(z/w)}{\mathcal{F}_{-m}(z/w)} \, \frac{\mathcal{F}_{-m}((-p^{*\frac{1}{2}})^{s}z/w)}{\mathcal{F}^*_{n}((-p^{*\frac{1}{2}})^{s}z/w)} \; t_{rs}(w) \, t_{mn}(z)
\label{eq:exchttmnrs}
\end{equation}
Only when $s=n$ does the exchange function identify with $\mathcal{Y}_{mn}(z/w)$, due to the identification $\mathcal{F}_{m}((-p^{*\frac{1}{2}})^{n}x)=\mathcal{F}^*_{-m}(x)^{-1}$. Similar results hold for the $t^*_{-n,-m}(z)$ generators.
\end{rmk}

\begin{cor}[``Localized'' extensions of the center]
\label{thmcommut}\ \\
We consider the surface $\mathscr{S}_{m,-m}$ where $m$ is \emph{odd} with $|m| \ne 1$, and impose the supplementary relations
\beq
c=-\dfrac{2}{m} \mb{and} -p^{\frac{1}{2}}=q^{2\lambda/m}, 
\eeq
where $\lambda$ is an integer, $\lambda \ne |m|-1$, such that $(\lambda,m)$ are coprimes with B\'ezout coefficients $(\beta,\beta')$, \emph{i.e.} $\beta\lambda+\beta'm=1$, and $(\beta+1,m)$ are coprimes.

Then the generators 
$t_{m,-m}(z)$ commute with the $L$ generators of the elliptic quantum algebra $\gelp$. 
\end{cor}
The proof of this corollary is given in section \ref{sect:proof-coro}. The characterization as ``localized'' reflects the fact that this extended center only exists on a submanifold of the surface $\mathscr{S}_{m,-m}$.

\begin{cor}[Scaling limit]\ \\
Defining the scaling limit as
$p=1+\eps$ {and} $q=1+\eta\,\eps$ {with} $\eps\to0$, we observe that
\bea
g^{(k)}(z)&=&1-\eps^2\,\frac{k(k+2\eta)}2 \frac{z}{(1-z)^2} +o(\eps^2) 
\\
g^{*(k)}(z)&=&1-\eps^2\,\frac{k(1-2\eta c)\big(k(1-2\eta c)+2\eta\big)}2 \frac{z}{(1-z)^2} +o(\eps^2) 
\eea
leading to
\begin{equation}
g_{mn}(z) = 1-\eps^2\Big\{  \beta_m-(1-2\eta c)^2\beta_n-2\eta^2 c(1-2\eta c)n(n-2)
\Big\}\frac{z}{(1-z)^2} +o(\eps^2) 
\end{equation}
where
\begin{equation}
\beta_\ell = \frac{|\ell|(|\ell|-1)(2|\ell|-1)}6+\eta\ell(\ell-2).
\end{equation}
\end{cor}
All $g_{mn}$ functions have the same $z$-dependence, which coincides (up to a surface dependent coefficient) with the scaling limit of the structure function in \cite{SKAO}. 
This scaling limit yielded the undeformed Virasoro algebra. This leads us to characterize our algebraic structures as deformed Virasoro algebras.

Note that the form of the exchange relations \eqref{eq:exchtt} to \eqref{eq:exchttstar} suggest the existence of a possible connection between the $t_{mn}$ and $t^*_{-n,-m}$ generators. Indeed, one obtains:
\begin{prop}[Relation between $t_{mn}(z)$ and $t_{-n,-m}^*(z)$]
\label{prop:ttstar-qdet}\ \\
On the surface $\mathscr{S}_{mn}$, we have the following relation 
\begin{equation}
t_{-n,-m}^*(z) = \frac{\qdet L(qz)}{\qdet L((-p^{\frac{1}{2}})^{-m}z)} \; t_{mn}(qz) \,.
\end{equation}
 $\qdet L(z)$ is the quantum determinant for the elliptic quantum algebra $\gelp$, see Appendix \ref{sect:qdet}.
\end{prop}
\proof Direct calculation. \qed

It is thus sufficient to study the generators $t_{mn}(z)$. We now establish sufficient conditions for  the subalgebra generated by $t_{mn}(z)$ to be abelian.

\begin{prop}[Abelian subalgebras in $\gelp$]
\label{thmtabel}\ \\
On the surface $\mathscr{S}_{mn}$, the generators $t_{mn}(z)$ realize an abelian subalgebra in $\gelp$ when one of the following conditions is satisfied (with here $N=2$): 
\begin{itemize}

\item for $|m|,|n|>1$: 
\begin{equation}\label{eq:abel1}
c = \frac{N}{nm} \big( \lambda' m-\lambda n \big) 
\;, \;\; -p^{\frac{1}{2}} = q^{-N\lambda/m} \;, \;\; -p^{*\frac{1}{2}} = q^{-N\lambda'/n}
\end{equation}
where $\lambda,\lambda'\in\ZZ\setminus\{0\}$ and $\lambda+\lambda'=1$. 

\item for $|n|=1, |m|>1$:
\begin{equation}\label{eq:abel2}
c=Nn\big(1-\lambda(m+n)\big) 
\;, \;\; -p^{\frac{1}{2}} = q^{-N\lambda} \;, \;\; -p^{*\frac{1}{2}} = q^{-Nn(1-\lambda m)}
\end{equation}
where $\lambda \in \ZZ/2$ or $\lambda \in \ZZ/u$, $u$ being any divisor of $m$ or $m+n$. 

\item for $|m|=1,|n|>1$:
\begin{equation}\label{eq:abel3}
c=Nm\big(\lambda'(n+m)-1\big) 
\;, \;\; -p^{\frac{1}{2}} = q^{-Nm(1-\lambda' n)} \;, \;\; -p^{*\frac{1}{2}} = q^{-N\lambda'}
\end{equation}
where $\lambda' \in \ZZ/2$ or $\lambda' \in \ZZ/u'$, $u'$ being any divisor of $n$ or $n+m$.

\item for $m=n=\pm 1$, formulas \eqref{eq:abel2}--\eqref{eq:abel3} also hold with $\lambda,\lambda' \in \ZZ/2$ 
and $\lambda+\lambda'=1$.
\item for $m+n=0$ with $n>0$ and odd:
\beq\label{eq:abel4}
c=\frac{N}{n}\;, \  -p^{\frac{1}{2}} = q^{-\frac{n-1}{2n}N}\;, \  -p^{*\frac{1}{2}} = q^{-\frac{n+1}{2n}N}.
\eeq
\end{itemize}
\end{prop}
The proof is given in section \ref{subsect:C2}. Once we get an abelian subalgebra, a Poisson structure can be defined, 
see section \ref{sect:poisson}.

\begin{prop}[Elliptic Liouville formula]\label{prop:liouville}\ \\
The generator $t_{0,2}(z) = \half\tr \big( L(q^{-2}z) L(z)^{-1} \big)$ lies in the center of the elliptic quantum algebra $\gelp$. 
This Liouville-type generator is related to the quantum determinant through:
\begin{equation}
t_{0,2}(z) = \frac{\qdet L(q^{-1}z)}{\qdet L(z)} \,.
\end{equation}
\end{prop}
\proof Direct calculation using relations \eqref{eq:comat} and \eqref{eq:lchapeau}. \qed
Note that this last statement does not depend on any choice of surface. It relies only on the properties of the quantum determinant.

\subsubsection{Case of the \ellipt{2} algebra}
The construction presented in section \ref{sect:t-n=2} can be readily repeated using the algebra defined by relations \eqref{eq:elpalt1}-\eqref{eq:elpalt2}, mimicking however expressions of the $t_{mn}$ or $t_{-n,-m}^*$ generators in terms of $L^\pm$ instead of $L$. We therefore introduce the following generators:
\begin{align}
& \ft_{mn}(z) = \tr \Big( (g^{\frac{1}{2}} h g^{\frac{1}{2}})^{-m+1}\, \cL^+\big((-p^{*\frac{1}{2}})^{n+1}q^{\frac{c}{2}}z\big)\,
 (g^{\frac{1}{2}} h g^{\frac{1}{2}})^{-n-1}\, \cL^-(z)^{-1} \Big), \\
& \ft_{-n,-m}^*(z) = \tr \Big( (g^{\frac{1}{2}} h g^{\frac{1}{2}})^{n+1}\, \cL^-\big((-p^{\frac{1}{2}})^{-m-1}z\big)^{-1}\,
 (g^{\frac{1}{2}} h g^{\frac{1}{2}})^{m+1}\, \cL^+\big(q^{-\frac{c}{2}}z\big) \Big) .
\end{align}
It is obvious that the surface on which they close quadratically is still $\mathscr{S}_{mn}$, since the structure constants in \eqref{eq:elpalt1}-\eqref{eq:elpalt2} differ only by scalars from $\gelp$.

We focus here on a very specific set of generators. It turns out that the above expressions become particularly simple in the following cases:
\begin{align}
\ft_{m,-1}(z) &= \tr \big( (g^{\frac{1}{2}} h g^{\frac{1}{2}})^{-m+1} \cL^+(q^{\frac{c}{2}}z) \cL^-(z)^{-1} \big), \\
\ft_{-n,1}^*(z) &= \tr \big( (g^{\frac{1}{2}} h g^{\frac{1}{2}})^{n+1} \cL^-(z)^{-1} \cL^+(q^{-\frac{c}{2}}z) \big) .
\end{align}
The exchange relations for the $\ft_{m,-1}(z)$ or $\ft_{-n,1}^*(z)$ generators on the suitable surfaces take the form
\begin{align}\label{Y-tilde}
\ft_{m,-1}(z)\,\ft_{m,-1}(w) &=\wt{\mathcal{Y}}_{m,-1}(z/w) \, \ft_{m,-1}(w) \, \ft_{m,-1}(z), \\
\ft_{-n,1}^*(z)\,\ft_{-n,1}^*(w) &=\wt{\mathcal{Y}}_{-n,1}^*(z/w) \, \ft_{-n,1}^*(w) \, \ft_{-n,1}^*(z) ,
\end{align}
where the new structure functions $\wt{\mathcal{Y}}_{m,-1}$ and $\wt{\mathcal{Y}}_{-n,1}^*$ read
\begin{align}
\wt{\mathcal{Y}}_{m,-1}(x) &= \frac{\tau_2(q^{\frac{1}{2}-c}x)\tau_2(q^{\frac{1}{2}}x^{-1})}{\tau_2(q^{\frac{1}{2}}x)\tau_2(q^{\frac{1}{2}-c}x^{-1})} \; {\mathcal{Y}}_{m,-1}(x), \\
\wt{\mathcal{Y}}_{-n,1}^*(x) &= \frac{\tau_2(q^{\frac{1}{2}+c}x^{-1})\tau_2(q^{\frac{1}{2}}x)}{\tau_2(q^{\frac{1}{2}}x^{-1})\tau_2(q^{\frac{1}{2}+c}x)} \; {\mathcal{Y}}_{-1,n}(x).
\end{align}
\begin{prop}[Deformed Virasoro algebra]\ \\
The structure function $\wt{\mathcal{Y}}_{2,-1}(x)$ reduces to the structure function of the DVA obtained in \cite{AKOS9612233}:
\beq
\wt{\mathcal{Y}}_{2,-1}(x)=\frac{g^{(1)}(x^2)}{g^{(1)}(x^{-2})} 
\mb{with} g^{(1)}(z)= \exp\Big(\sum_{\ell=1}^\infty \frac{(1-p^{-\ell})(1-(pq^2)^\ell)}{1+q^{2\ell}}\, \frac{z^{\ell}}{\ell}\Big).
\eeq
In the same way, $\wt{\mathcal{Y}}^*_{-2,1}(x)$ also reproduces the structure constant of \cite{AKOS9612233}. 
\end{prop}
\proof Direct calculation starting from \eqref{Y-tilde} and using \eqref{Y:gg}-\eqref{def:g}.\qed
The correspondence with the notation used in \cite{AKOS9612233} is as follows
\beq
Q=p^{-1} \mb{,} P=q^2 \mb{and} t^{-1}=\frac PQ=pq^2
\eeq
where $P$, $Q$ and $t$ are the parameters used in \cite{AKOS9612233}.

Hence, the generators $\ft_{2,-1}(z)$ and $\ft_{-2,1}^*(z)$ provide an algebraic construction for the deformed Virasoro algebra 
obtained in \cite{AKOS9612233}, using a totally different approach. 

We have in this way established the deep consistency between the construction of the DVA by vertex operators built from deformed bosons advocated in \cite{SKAO} and its alternative construction as a subalgebra of an elliptic algebra (related but not equivalent to $\elp$ built from bilinear functionals of the generating quantum Lax matrix. 
This consistency was lacking until now: the exchange factor for bilinear functionals of the elliptic Lax operators,  closest to the exact DVA, were in fact the \emph{square} of the structure function of \cite{SKAO}, see \cite{AFRS99}. 
We believe this is a major step in the understanding of the connection between DVA (and later DWA) and the elliptic algebras. 
Of course this begs the question of the connection between the \ellipt{2} algebra used here, and the canonical elliptic quantum algebra.

Other questions arise: it must be noted that MacDonald polynomials are natural eigenfunctions of the generating operators, as a direct consequence of the $q$-bosonization technique of vertex operators, used to construct the deformed Virasoro algebra \cite{AKOS9612233}. It would be interesting to investigate whether some representation, on the same line as $q$-bosonization, may be available for our construction starting from $L$ operators.

The next question would be to see whether any such construction may be implemented for the distinct algebra structures corresponding to the other surfaces. More generally, the search of orthogonal polynomials, eigenfunctions of the deformed Virasoro generators $t_{mn}(z)$, $t_{mn}^*(z)$, $\ft_{mn}(z)$ or $\ft_{mn}^*(z)$ is a appealing question, but is beyond the scope of the present paper.

\subsection{Generalization to elliptic algebras based on $gl(N)$\label{sect:exchange}}
We focus here on the algebra $\gelpn$ but obviously the same type of generalization can be done for the algebra 
\ellipt{N}. We present only the modifications to the theorems that still apply and quote the ones that are not proved.
We also make some remarks that apply to the general $\gelpn$ case.
\subsubsection{Quadratic subalgebras in   $\gelpn$}
The theorem \ref{thmone}  now applies to surfaces
\beq
\mathscr{S}_{mn}\ :\ (-p^{\frac{1}{2}})^{m} (-p^{*\frac{1}{2}})^{n} = q^{-N}\,,\quad \forall m,n \in \ZZ,\quad n \ne 0.
\eeq
The structure constants $\cF$  have the following form in term of Jacobi $\Theta$ functions (for $m>0$):
\bea
\cF_m(x) &=& q^{m(2/N-2)}\,\prod_{k=0}^{m-1}\frac{\Theta_{q^{2N}}(q^2\,p^k\,x^2) \, \Theta_{q^{2N}}(q^2\,p^{-k}\,x^{-2})} {\Theta_{q^{2N}}(p^k\,x^2)\,\Theta_{q^{2N}}(p^{-k}\,x^{-2})}
\label{F:theta}
\\
\cF_{-m}(x) &=& q^{-m(2/N-2)}\,\prod_{k=1}^{m}\frac{\Theta_{q^{2N}}(p^{-k}\,x^2)\,\Theta_{q^{2N}}(p^{k}\,x^{-2})}{\Theta_{q^{2N}}(q^2\,p^{-k}\,x^2) \, \Theta_{q^{2N}}(q^2\,p^{k}\,x^{-2})}.
\label{F:theta2}
\eea
As far as $\cY$ functions are concerned they read
\beq\label{Y:FF}
\mathcal{Y}_{mn}(x) = \frac{\mathcal{F}^*_n(x) \mathcal{F}^*_{-n}(x)}{\mathcal{F}_m(x) \mathcal{F}_{-m}(x)} 
= \mathcal{Y}_{-m,-n}(x).
\eeq
This expression is valid for any $N$, while the factorized form \eqref{Y:gg} holds only for $N=2$.
\begin{rmk}\label{rmk:31}
Due to the expression of the matrices $g$ and $h$, one has $(g^{\frac{1}{2}} h g^{\frac{1}{2}})^{N} = \II$. Hence, the theorem 4 of ref. \cite{AFRS99} appears as a specific case of theorem \ref{thmone} when $m-1$ is a multiple of $N$ and $n=-1$. Indeed the possibility of any extra matrix terms $(g^{\frac{1}{2}} h g^{\frac{1}{2}})^{\#}$ in \eqref{eq:deftmn} was disregarded at the time, a key omission in the discussion.
\end{rmk}

\subsubsection{``Localized'' extensions of the center of $\gelpn$}
Corollary \ref{thmcommut} also extends to the $\gelpn$ algebra with now
\beq
c=-\dfrac{N}{m} \mb{and} -p^{\frac{1}{2}}=q^{N\lambda/m}, 
\eeq
and the same characterization of $\lambda$.

\begin{rmk}\label{rmk:31.5}
Note that unlike the $|m|=1$ case, the generators $t_{m,-m}(z)$ of corollary \ref{thmcommut} do not span an extended center of $\gelpn$. Indeed, one needs a supplementary relation between $p$ and $q$ to get the commutation property.
\end{rmk}

\subsubsection{Abelian subalgebras in $\gelpn$}
Again, the proposition \ref{thmtabel}  applies to the $\gelpn$ algebra, with now $N$ generic.
\begin{rmk}[Critical levels $c=\pm N$]
In the case $m=1$ and $n=-1$, the surface relation leads to fixing the central charge to $c=-N$ (critical level) and the generator $t_{mn}(z)$ commutes with $L(w)$.
This result was proven in \cite{AFRS99}.
In addition to the critical level $c=-N$, one also gets an extended center at $c=N$ corresponding to the case $m=-1$ and $n=1$.
This second critical level is specific to the elliptic case, and was not observed for quantum groups.
\end{rmk}
\begin{rmk}
It is interesting to notice that the generators $t_{0n}(z)$ and $t_{0,-m}^*(z)$ realize abelian subalgebras on the surfaces $\mathscr{S}_{0n}$ and $\mathscr{S}_{m0}$ respectively in $\gelpn$. This follows immediately from a direct calculation of \eqref{Y:FF}, taking into account the surface condition and the $q^N$-periodicity of $\mathcal{U}(x)$. 
This is a weaker statement than proposition \ref{prop:liouville} since it is only valid on the surfaces $\mathscr{S}_{0n}$ and $\mathscr{S}_{m0}$.
\end{rmk}

\subsubsection{Relation between $t_{mn}(z)$ and $t_{-n,-m}^*(z)$ and elliptic Liouville formula} 
Propositions \ref{prop:ttstar-qdet} and \ref{prop:liouville} do not apply to $\gelpn$.
The main reason is the lack of an expression for a quantum determinant when $N>2$. 
They are however replaced by the following (weaker) proposition:
\begin{prop}
On the surface $\mathscr{S}_{mn}$, the product of generators $t_{mn}(z) t^*_{-n,-m}(z)$ lies in the center of $\gelpn$.
\end{prop}
\proof
It is an obvious consequence of relations \eqref{eq:exchtL} and \eqref{eq:exchtLalt}.\qed
This proposition justifies that one focuses on $t_{mn}(z)$ only, at least when discussing the purely structural aspects of the algebra.

\section{Poisson structures\label{sect:poisson}}

Let us now focus again on the structures derived from the canonical elliptic quantum algebra $\gelpn$.
The results of the section \ref{sect:exchange} allow us to define Poisson structures on the corresponding abelian subalgebras in $\gelpn$.
The explicit construction of these Poisson structures follows the lines of \cite{AFRS99}. More precisely, 
when $c$ satisfies the conditions of proposition \ref{thmtabel} on the surface $\mathscr{S}_{mn}$, setting $p^{1-\epsilon} = q^{\alpha N \ell}$ where $\alpha$ depends on the considered case and $\ell\in\ZZ$, one defines a Poisson structure by
\begin{equation}
\big\{ t(z),t(w) \big\}_{\ell} = \lim_{\epsilon \to 0} \frac{1}{\epsilon} \big( t(z) t(w) - t(w) t(z) \big)
\label{eq:defpoisson}
\end{equation}
One gets the following theorem:
\begin{thm}
\label{thmpoisson}
The Poisson structure \eqref{eq:defpoisson} has the following expression:
\begin{equation}
\big\{ t(z),t(w) \big\}_{\ell} = f_{\ell}(z/w) \, t(z) t(w)
\end{equation}
where 
\begin{equation}\label{def:f}
f_{\ell}(x) = 2N\ell(\ln q) \big( 2I(x) - I(qx) - I(q^{-1}x) - (x \leftrightarrow x^{-1}) \big)
\end{equation}
$I(x)$ is given by the following expressions depending on the different cases of proposition \ref{thmtabel}.

$\bullet$ for $|m|,|n|>1$ ($\alpha=2/m$): 
\begin{align}
I(x) &= \sum_{s=1}^\infty \left( w\,\frac{x^2 q^{2Nsw/m}}{1-x^2 q^{2Nsw/m}} + w'\,\frac{x^2 q^{2Nsw'/n}}{1-x^2 q^{2Nsw'/n}} \right) + \half\,(w+w')\,\frac{x^2}{(1-x^2)} 
\end{align}
where $w=gcd(\ell,m)$ and $w'=gcd(\ell',n)$ ($w$ and $w'$ being taken positive). 

$\bullet$ for $|n|=1,|m|>1$ ($\alpha=1$) and $\ell$ even: 
\begin{equation}
I(x) = \half\,m(m+1) \Bigg[ \sum_{s=1}^\infty \frac{x^2 q^{2Ns}}{1-x^2 q^{2Ns}} + \half\,\frac{x^2}{1-x^2} \Bigg]
\label{eq:poisson2}
\end{equation}
while for $\ell$ odd 
\begin{equation}
I(x) = \left\lfloor \sfrac{m}{2} \right\rfloor \big(\left\lfloor \sfrac{m}{2} \right\rfloor + 1\big) \left( \sum_{s=1}^\infty \frac{x^2 q^{2Ns}}{1-x^2 q^{2Ns}} + \half\,\frac{x^2}{1-x^2} \right) + \left\lfloor \sfrac{m+1}{2} \right\rfloor^2 \sum_{s=0}^\infty \frac{x^2 q^{N(2s+1)}}{1-x^2 q^{N(2s+1)}} 
\label{eq:poisson3}
\end{equation}
$\lfloor x \rfloor$ is the floor function (integer part) of $x$. 

$\bullet$ for $|n|=1,|m|>1$ ($\alpha=2/u$ where $u$ is any divisor of $k_{\sup}$ or $k_{\sup}+1$ and $g\in\ZZ_{>0}$ is defined as $k_{\sup}/u$ or $(k_{\sup}+1)/u$ respectively): 
\begin{equation}
I(x) = g \Bigg[ \eta\sum_{s=1}^\infty \left( \frac{x^2 q^{2Ns}}{1-x^2 q^{2Ns}} + gw\,\frac{x^2 q^{2Nsw/u}}{1-x^2 q^{2Nsw/u}} \right) + \half\,(gw+\eta)\,\frac{x^2}{1-x^2} \Bigg]
\label{eq:poisson4}
\end{equation}
where $\eta=1$ when $gu=k_{\sup}$ and $\eta=-1$ when $gu=k_{\sup}+1$, $w=gcd(\ell,u)$, see the definition of $k_{\sup}$ after formula \eqref{eq:Ymn-alt}.

$\bullet$ for $|m|=1,|n|>1$: the formulas are analogous to \eqref{eq:poisson2}-\eqref{eq:poisson4} ($\ell$ being replaced by $\ell'$, $k_{\sup}$ by $k'_{\sup}$ and $m$ by $n$), up to non relevant normalization factors that can be absorbed by a redefinition of $\epsilon$.
\end{thm}
\proof
One has by definition of the Poisson structure 
\begin{align}
\big\{ t(z),t(w) \big\}_{\ell} &= \lim_{\epsilon \to 0} \frac{1}{\epsilon} \big( t(z) t(w) - t(w) t(z) \big) \nonumber \\
&= -\left.\frac{d\mathcal{Y}_{mn}}{d\epsilon}(z/w)\right\vert_{\epsilon=0} t(z) t(w) = -\left.\frac{d\ln\mathcal{Y}_n}{d\epsilon}(z/w)\right\vert_{\epsilon=0} t(z) t(w)
\end{align}
since $\mathcal{Y}_{mn}(x) = 1$ for $\epsilon = 0$, $p$ and $q$ being linked by the proper relation. \\
The explicit form of the Poisson structure is given by a direct (somewhat lengthy) calculation of the derivative and the definition \eqref{theta:prod} of the Jacobi $\Theta$ functions as absolute convergent products for $|q|<1$.
\qed

Note that in particular \eqref{eq:poisson2} reproduces (up to a  normalization) the exact structure of the classical DVA in \cite{FF,SKAO}. However its natural quantization \eqref{eq:exchtt} does not, as already commented.

\section{Technical proofs\label{sect:proof}}
\subsection{Proof of Theorem \ref{thmone} \label{subsect:C1}}

The proof is given for the general case of $\gelpn$ and thus applies to both theorem \ref{thmone} and its generalization presented in section \ref{sect:exchange}.

We start by considering the following operator
\begin{equation}
t(z) = \tr \big( M L(\alpha z) \tilde M L(z)^{-1} \big) \equiv \tr Q(z)
\label{eq:t}
\end{equation}
where the matrices $M$ and $\tilde M$ will be characterized later. One has
\begin{align}
t(z) L_1(w)^{-1} = \tr_0 \big( M_0 L_0(\alpha z) \tilde M_0 {L_0(z)}^{-1} L_1(w)^{-1} \big) \,.
\label{eq:tLm}
\end{align}
By the $RLL$ relation \eqref{eq232} one gets
\begin{equation}
t(z) L_1(w)^{-1} = \tr_0 \Big( M_0 L_0(\alpha z) \tilde M_0 \widehat R_{01}^{*}(z/w) L_1(w)^{-1} L_0(z)^{-1} \, \widehat R_{01}^{-1}(z/w) \Big) \,.
\end{equation}
Assuming that the matrix $\tilde M_0$ obeys the relation
\begin{equation}
\tilde M_0 \widehat R_{01}^{*}(x) = {\mathcal{F}^*(x)} \widehat R_{01}^{*}(\alpha x) \tilde M_0
\label{eq:echN}
\end{equation}
for some function ${\mathcal{F}^*(x)}$ to be determined, one gets
\begin{align}
t(z) L_1(w)^{-1} &= {\mathcal{F}^*(z/w)} \, \tr_0 \Big( M_0 L_0(\alpha z) \widehat R_{01}^{*}(\alpha z/w) L_1(w)^{-1} \tilde M_0 L_0(z)^{-1} \, \widehat R_{01}^{-1}(z/w) \Big) \nonumber \\
&= \mathcal{F}^*(z/w) \, \tr_0 \Big( M_0 L_1(w)^{-1} \widehat R_{01}(\alpha z/w) L_0(\alpha z) \tilde M_0 L_0(z)^{-1} \, \widehat R_{01}^{-1}(z/w) \Big) \nonumber \\
&= \mathcal{F}^*(z/w) \, L_1(w)^{-1} \tr_0 \Big( M_0 \widehat R_{01}(\alpha z/w) L_0(\alpha z) \tilde M_0 L_0(z)^{-1} \, \widehat R_{01}^{-1}(z/w) \Big) \,.
\end{align}
Now assuming that the matrix $M$ obeys a relation similar to \eqref{eq:echN}, namely
\begin{equation}
M_0 \widehat R_{01}(x) = {\mathcal{F}(x)} \widehat R_{01}(\beta x) M_0 \,,
\label{eq:echM}
\end{equation}
one obtains
\begin{align}
t(z) L_1(w)^{-1} &= {\mathcal{F}^*(z/w)} \, {\mathcal{F}(\alpha z/w)} \, L_1(w)^{-1} \tr_0 \Big( \widehat R_{01}(\alpha \beta z/w) M_0 L_0(\alpha z) \tilde M_0 L_0(z)^{-1} \, \widehat R_{01}^{-1}(z/w) \Big) \nonumber \\
&= {\mathcal{F}^*(z/w)} \, {\mathcal{F}(\alpha z/w)} \, L_1(w)^{-1} \tr_0 \Big( \widehat R_{01}(\alpha \beta z/w) Q_0(z) \widehat R_{01}^{-1}(z/w) \Big).
\label{eq:310}
\end{align}
Note that one has $\tr_0 \big( \widehat R_{01} Q_0 \widehat R_{01}^{-1} \big) = \tr_0 \big( Q_0 (\widehat R_{01}^{-1})^{t_1} (\widehat R_{01})^{t_1} \big)^{t_1}$. Therefore a trivial dependence in space 1 is left under the trace over space 0 in \eqref{eq:310} when $\big(\widehat R_{01}^{-1}(x)\big)^{t_1} \big(\widehat R_{01}(\alpha \beta x)\big)^{t_1}$ is proportional to the unit matrix, i.e.
\begin{equation}
\widehat R_{01}(\alpha \beta x) \propto \Big(\big((\widehat R_{01}^{-1}(x))^{t_1}\big)^{-1}\Big)^{t_1} = \widehat R_{01}(q^{-N}x)
\end{equation}
which is satisfied when $\alpha\beta=q^{-N}$ and the proportionality coefficient is one.

We come now to the solutions of equations \eqref{eq:echN} and \eqref{eq:echM}, which connect the matrices $\widehat R$ for different arguments. The crucial observation is to notice that this connection can be performed using two equations, namely the quasi-periodicity property \eqref{eq225} and the antisymmetry property \eqref{eq224}. Postponing the use of the antisymmetry property to the remark \ref{rmk:antisym}, it follows that a first set of possible solutions of \eqref{eq:echN} and \eqref{eq:echM} is given by
\begin{equation}
M = (g^{\frac{1}{2}} h g^{\frac{1}{2}})^{-m} \;\; \text{and} \;\; \beta = (-p^{\frac{1}{2}})^{m} \quad , \quad
\tilde M = (g^{\frac{1}{2}} h g^{\frac{1}{2}})^{-n} \;\; \text{and} \;\; \alpha = (-p^{*\frac{1}{2}})^{n}
\end{equation}
where $m,n \in \ZZ$. Given the values of $\alpha$ and $\beta$, one gets the exchange relation (now expliciting for the operator $t(z)$ as well as the functions $\mathcal{F}$ and $\mathcal{F}^*$ the indices $m,n$ on which they depend)
\begin{equation}
t_{mn}(z) \, L(w)^{-1} = \mathcal{F}^*_n(z/w) \, \mathcal{F}_m((-p^{*\frac{1}{2}})^{n}z/w) \, L(w)^{-1} \, t_{mn}(z)
\end{equation}
provided the parameters  $q$, $p$, $c$ lie on the surface $\mathscr{S}_{mn}$ defined by the relation 
\begin{equation}
(-p^{\frac{1}{2}})^{m} (-p^{*\frac{1}{2}})^{n} = q^{-N} \,.
\end{equation}
The function $\mathcal{F}_a(x)$ ($a\in\ZZ$) is then given by \eqref{eq:exprFn} thanks to \eqref{eq225} and \eqref{eq:unitarity}, and $\mathcal{F}^*_a(x) = \mathcal{F}_a(x)\big\vert_{p \to p^*}$.  The surface condition allows us to rewrite $\mathcal{F}_m((-p^{*\frac{1}{2}})^{n}x)$ as $\mathcal{F}_{-m}(x)^{-1}$ and we  get the exchange relation \eqref{eq:exchtL}.
Using the expression \eqref{def:U}, we get the expressions \eqref{F:theta}, \eqref{F:theta2} and \eqref{Y:FF}.

The proof of relation \eqref{eq:exchtLalt} follows the same lines.
The relations \eqref{eq:exchtt}, \eqref{eq:ttstar} and \eqref{eq:exchttstar} are direct consequences of relations \eqref{eq:exchtL} and \eqref{eq:exchtLalt}.

In the case $N=2$, the Jacobi $\Theta$ functions have period $q^4$, which allows to simplify expressions involving $q^2$ identified as a half-period, see relation \eqref{id:theta}. Expanding the Jacobi $\Theta$ functions as infinite products, see \eqref{theta:prod}, yields the factorization \eqref{Y:gg} of the structure functions $\cY_{mn}(z)$ with
\beq\label{g:prod}
g^{(k)}(z)= \frac{1-z}{1-p^kz}\frac{(p^kz;q^4)\,(q^2p^{-k}z;q^4)}{(p^{-k}z;q^4)\,(q^2p^{k}z;q^4)}.
\eeq
Finally, \eqref{def:g} is a direct consequence of \eqref{g:prod}.
\qed
\begin{rmk}\label{rmk:antisym}
Exploiting the antisymmetry property \eqref{eq224}, we get another solution of \eqref{eq:echN} and \eqref{eq:echM} given by ($a,b$ being integers)
\begin{equation}
\begin{split}
& M = g^b(g^{\frac{1}{2}} h g^{\frac{1}{2}})^{-m} \;\; \text{and} \;\; \beta = (-1)^b(-p^{\frac{1}{2}})^{m} \\
& \tilde M = g^a(g^{\frac{1}{2}} h g^{\frac{1}{2}})^{-n} \;\; \text{and} \;\; \alpha = (-1)^a(-p^{*\frac{1}{2}})^{n}
\end{split}
\end{equation}
The surface $\tilde{\mathscr{S}}_{mn}$ to be considered is now defined by the relation $(-p^{\frac{1}{2}})^{m} (-p^{*\frac{1}{2}})^{n} = (-1)^{a+b}q^{-N}$. The extra factor $(-1)^{a+b}$ can then be interpreted as a (complex) shift in the central charge: $c \to c - \dfrac{\ln\varphi_{ab}}{\ln q}$ where the phase $\varphi_{ab}$ satisfies $(\varphi_{ab})^n=(-1)^{a+b}$.
The effect of the new conditions on the exchange functions is also to multiply them by a phase: the exchange functions in \eqref{eq:exchtL} become $\widetilde{\mathcal{F}}_m(x) = \omega^{2b} \mathcal{F}_m(x)$ and $\widetilde{\mathcal{F}}^*_n(x) = \omega^{2a} \mathcal{F}^*_n(x)$. 
We leave to the reader the working out of the suitable changes for the exchange functions $\mathcal{Y}_{mn}$ between the $t_{mn}(z)$ (as well as $t^*_{mn}(z)$) operators, and the abelianity conditions.
\end{rmk}

\subsection{Proof of corollary \ref{thmcommut}\label{sect:proof-coro}}
Again, the proof is given for the general case of $\gelpn$ and thus applies to both proposition \ref{thmtabel} and its generalization presented in section \ref{sect:exchange}.

Given \eqref{eq:exchtL}, a commutative exchange relation can be considered in the case $m+n=0$, hence $c=-N/m=N/n$. We suppose here that $|m| \ne 1$, see remark \ref{rmk:31}. The exchange function is then given by
\begin{equation}
\prod_{k=1}^m \; \frac{\mathcal{U}\big((-p^{*\frac{1}{2}})^{-k} x\big)}{\mathcal{U}\big((-p^{\frac{1}{2}})^{-k} x\big)} \quad \text{for $m>0$} \qquad \text{and} \qquad \prod_{k=0}^{n-1} \; \frac{\mathcal{U}\big((-p^{\frac{1}{2}})^{k} x\big)}{\mathcal{U}\big((-p^{*\frac{1}{2}})^{k} x\big)} \quad \text{for $n>0$}.
\label{eq:ratio}
\end{equation}
Consider first the case $m>0$. The ratio \eqref{eq:ratio} is equal to 1 if each term indexed by $k$ in the numerator simplifies with the term indexed by $\sigma(k)$ in the denominator where $\sigma \in \mathfrak{S}_m$, up to a power of $q^N$ since the function $\mathcal{U}$ is $q^N$-periodic. Hence, one gets $\ell(k) - \dfrac{k}{m} = \alpha\big(k-\sigma(k)\big)$, $\ell(k)$ being an integer depending on $k$ while $\alpha$ is constant. 

Looking for solutions of the type $\alpha=\lambda/m$ where $(\lambda,m)$ are coprimes, one obtains 
\beq
m\big(\ell(k+1)-\ell(k)\big) + \lambda\big(\sigma(k+1)-\sigma(k)-1\big)=1
\label{eq:(1)}
\eeq
 with the boundary equation 
\beq
m\ell(1) + \lambda\big(\sigma(1)-1\big)=1.
\label{eq:(2)}
\eeq
Eqs. \eqref{eq:(1)}-\eqref{eq:(2)} cannot be realized with the choice $\lambda=m-1$. Restricting to $\lambda \ne m-1$, let $(\beta,\beta')$ be the B\'ezout coefficients of $(\lambda,m)$. The solution of \eqref{eq:(1)} is given by $\ell(k+1)-\ell(k)=\beta'$ and $\sigma(k+1)-\sigma(k)-1=\beta \mod m$, while \eqref{eq:(2)} leads to $\sigma(1)=m+\beta+1$. When $m$ is even, $\beta$ has to be odd, hence it is impossible to generate odd values of $\sigma(k)$. This leaves us only with odd values for $m$. One then gets $\sigma(k)=m+k(\beta+1)$. When $(\beta+1,m)$ are not coprimes, this leads to cycles in the set of values of $\sigma(k)$. Otherwise, $\sigma(k)=m+k(\beta+1)$ span the set of values $(1,...,m)$, which concludes the proof for $m>0$. 
 
The case $m<0$ runs along similar lines.

\subsection{Proof of Proposition \ref{thmtabel} \label{subsect:C2}}
Once more, the proof is given for the general case of $\gelpn$ and thus applies for both proposition \ref{thmtabel} and its generalization presented in section \ref{sect:exchange}.

Starting from expression \eqref{Y:FF}, we look for sufficient conditions for the function $\mathcal{Y}_{mn}(x)$ to be equal to 1. 
Using the expression of $\cF$, see eq. \eqref{eq:exprFn}, we get the following expression for $\mathcal{Y}_{mn}(x)$:
\beq
\mathcal{Y}_{mn}(x)= \frac{\displaystyle \prod_{k=1}^{|m|} \mathcal{U}\big((-p^{\frac{1}{2}})^{-k} x\big) 
\prod_{k'=0}^{|n|-1} \mathcal{U}\big((-p^{*\frac{1}{2}})^{k'} x\big)}
{\displaystyle \prod_{k=0}^{|m|-1} \mathcal{U}\big((-p^{\frac{1}{2}})^{k} x\big) 
\prod_{k'=1}^{|n|} \mathcal{U}\big((-p^{*\frac{1}{2}})^{-k'} x\big)}.
\eeq
Moreover, using the surface relations of $\mathscr{S}_{mn}$, it can be rewritten as
\begin{equation}
\mathcal{Y}_{mn}(x) = 
\prod_{k=k_{\inf}}^{k_{\sup}} \frac{\mathcal{U}\big((-p^{\frac{1}{2}})^{-k} x\big)}{\mathcal{U}\big((-p^{\frac{1}{2}})^{k} x\big)} \;
\prod_{k'=k'_{\inf}}^{k'_{\sup}} \frac{\mathcal{U}\big((-p^{*\frac{1}{2}})^{k'} x\big)}{\mathcal{U}\big((-p^{*\frac{1}{2}})^{-k'} x\big)}
\label{eq:Ymn-alt}
\end{equation}
where $k_{\sup}=|m|$, $k'_{\sup}=|n|-1$ or $k_{\sup}=|m|-1$, $k'_{\sup}=|n|$ when $mn > 0$ ; $k_{\sup}=|m|-1$, $k'_{\sup}=|n|-1$ when $mn < 0$ ; independently, $k_{\inf}$ and $k'_{\inf}$ can be chosen arbitrarily in $\{0,1\}$ since a zero value for $k_{\inf}$ or $k'_{\inf}$ does not contribute in \eqref{eq:Ymn-alt}.

Since the arguments of the $\mathcal{U}$ functions in \eqref{Y:FF} contain shifts in respectively $-p^{\frac{1}{2}}$ or $-p^{*\frac{1}{2}}$, we are led to demand, when $|m| \ne |n|$, that both product over $k$ and $k'$ be equal to 1, since it seems complicated to assume consistently in this case that ``cross-cancellations'' occur between $p$ and $p^*$-shifted terms. Let us start the proof by assuming that $|m| \ne |n|$. For the first product, this cancellation certainly occurs whenever each term indexed by $k$ in the numerator simplifies with the term indexed by $\sigma(k)$ in the denominator, where $\sigma$ is  some permutation of the $k_{\sup}-k_{\inf}+1$ indices of the product. 
The simplification occurs whenever the arguments match up to a power of $q^N$ due to the $q^N$-periodicity of $\mathcal{U}$. A similar argument holds for the second product with another permutation $\sigma'$ of the $k'_{\sup}-k'_{\inf}+1$ indices of the product.
Using the surface condition, one is led for each $k$ and each $k'$ to the relations
\begin{equation}
\begin{split}
(a) \qquad & N-nc = \frac{N\ell(n+m)}{k+\sigma(k)} \\
(b) \qquad & N+mc = \frac{N\ell'(n+m)}{k'+\sigma'(k')} 
\label{eq:320}
\end{split}
\end{equation}
where $\ell$ and $\ell'$ are integer numbers \emph{a priori} depending on $k$ and $k'$ respectively. \\
These two relations imply a compatibility condition by eliminating the central charge $c$:
\begin{equation}
m\,\frac{\ell}{k+\sigma(k)} + n\,\frac{\ell'}{k'+\sigma'(k')} = 1
\label{eq:321}
\end{equation}
Let us immediately note that when $|m|=1$ or $|n|=1$, one of the two products in \eqref{eq:Ymn-alt} is equal to 1. In that case, only one of the two above relations survives in \eqref{eq:320} and there is no compatibility condition \eqref{eq:321}.

Since $c$ has to be independent of $k$ and $k'$, \eqref{eq:320} leads to $(i)$ $\ell$ is proportional to $(k+\sigma(k))$ or $(ii)$ $k+\sigma(k)$ is constant. Similar conditions hold for $\ell'$, $k'$ and $\sigma'$. \\
In the case $(i)$, one has $\ell = \alpha\big(k+\sigma(k)\big)$. A careful analysis of the possible values of $\alpha$ shows that $\alpha \in \ZZ/2$ or $\alpha \in \ZZ/u$ where $u$ is any divisor of $k_{\sup}$ or $k_{\sup}+1$. \\
In the case $(ii)$, one sets $k+\sigma(k) = h = \text{constant}$. 
In the product \eqref{Y:FF} the factors $k=0$  in numerator and denominator cancel out. 
One may thus consider this product as running either from $0$ to $k_{\sup}$, or from $1$ to $k_{\sup}$, corresponding to the possible values of $k_{\inf}$.
The two choices of range for the running index $k$ however leads to two possible, nonequivalent choices for the canceling permutation $\sigma$, the corresponding constant $h$ being equal to $k_{\sup}+k_{\inf}$.

In the case $|n|=1$, one is left with $(a)$ in \eqref{eq:320}. Hence from the above discussion one gets $c=nN\big(1-\lambda(m+n)\big)$ where $\lambda \in \ZZ/2$ or $\lambda \in \ZZ/u$, $u$ being any divisor of $k_{\sup}$ or $k_{\sup}+1$. Similarly, when $|m|=1$, one is left with $(b)$ in \eqref{eq:320} solved by $c=mN\big(\lambda'(n+m)-1\big)$ where $\lambda' \in \ZZ/2$ or $\lambda' \in \ZZ/u'$, $u'$ being any divisor of $k'_{\sup}$ or $k'_{\sup}+1$. When $m=n=\pm 1$, it is easy to check that these results also hold.

We suppose now that $|m|$ and $|n|$ are different from 1. It follows that one has to examine four possibilities depending whether case $(i)$ or case $(ii)$ holds for each relation in \eqref{eq:320}.  
\begin{itemize}
\item[(a)] $k+\sigma(k) = h$ and $k'+\sigma'(k') = h'$ ($h,h'$ constants) \\
The compatibility condition \eqref{eq:321} reads $m\,\dfrac{\ell}{h}+n\,\dfrac{\ell'}{h'}=1$. 
Since $h$ is given by $k_{\sup}$ or $k_{\sup}+1$ and $k_{\sup} = |m|$ or $|m|-1$ (and similar relations for $h'$ with $k'_{\sup} = |n|-1$ or $|n|$), one ends up with the following cases for the choice of the pair of integers $(h,h')$: $(|m|,|n|)$, $(|m| \pm 1,|n|)$, $(|m|,|n| \pm 1)$, $(|m| \pm 1,|n| \mp 1)$ when $mn>0$ and $(|m|,|n|)$, $(|m|-1,|n|)$, $(|m|,|n|-1)$, $(|m|-1,|n|-1)$ when $mn<0$, from which it follows that one can restrict the discussion to $m,n>0$ since $\ell,\ell'\in\ZZ$. \\
When $(h,h')=(m,n)$, one has $\ell+\ell'=1$, hence the result where $\lambda=\ell$ is an arbitrary nonzero integer and $\lambda'=\ell'$. The other possibilities for $(h,h')$ only lead to subcases of the previous result. Indeed,
when $(h,h')=(m \pm 1,n)$, the choice $\ell=\beta(m \pm 1)$, $\ell'=1-\beta m$ with $\beta\in\ZZ$ solves the problem and one obtains \eqref{eq:abel1} with $\lambda=1-\ell'$.
Analogously, when $(h,h')=(m,n \pm 1)$, a solution is given by $\ell'=\beta'(n \pm 1)$, $\ell=1-\beta' n$ with $\beta'\in\ZZ$ and 
one gets \eqref{eq:abel1} with $\lambda=\ell$. When $(h,h')=(m \pm 1,n \mp 1)$, there is no solution if $\gcd(m,n) > 1$. 
If $(m,n)$ are coprimes, let $(\beta,\beta')$ denote their B\'ezout coefficients. Then $\ell=\beta(m \pm 1)$ and 
$\ell'=\beta'(n \mp 1)$ solves the problem and one gets \eqref{eq:abel1} with $\lambda=m\beta$ and $\lambda'=n\beta'$.
\item[(b)] $\ell = \alpha\big(k+\sigma(k)\big)$ and $\ell' = \alpha'\big(k'+\sigma'(k')\big)$ \\
The compatibility condition \eqref{eq:321} then reads $m\alpha+n\alpha'=1$. Let $g = \gcd(m,n)$ and $(\beta,\beta')$ be 
the B\'ezout coefficients for $(m,n)$, i.e. the integers such that $m\beta+n\beta'=g$. Note that $g$ is a divisor of $k_{\sup}$ 
and $k'_{\sup}+1$ or of $k_{\sup}+1$ and $k'_{\sup}$ depending on the alternative formula we start with for $\mathcal{Y}_{mn}(x)$. 
Then the choice $\alpha=\beta/g$ and $\alpha'=\beta'/g$ solves \eqref{eq:320}--\eqref{eq:321}, which is a subcase of the 
theorem with $\lambda=m\alpha$ and $\lambda'=n\alpha'$.
\item[(c)] $\ell = \alpha\big(k+\sigma(k)\big)$ and $k'+\sigma'(k') = h'$ \\
The compatibility condition \eqref{eq:321} is now $m\alpha+n\,\dfrac{\ell'}{h'}=1$. 
When $h'=|n|$, the choice $\alpha = \beta/u$ where $\beta\in\ZZ$ and $u$ is a divisor of $m$, $\ell'=\sgn(n)(1-\alpha m)$ is always possible. 
When $h'=|n| \pm 1$, let $g=\gcd(m,n)$. If $(m/g,n)$ are coprimes, there is a solution given by $\alpha=\beta/g$ and  $\ell'=\beta'h'$ where $(\beta,\beta')$ are B\'ezout coefficients of $(m/g,n)$. In both cases, one gets \eqref{eq:abel1} with $\lambda=m\alpha$.
\item[(d)] $k+\sigma(k) = h$ and $\ell' = \alpha'\big(k'+\sigma'(k')\big)$ \\
This case is similar to the previous one. The compatibility condition \eqref{eq:321} is here $m\,\dfrac{\ell}{h}+n\alpha'=1$. 
When $h=|m|$, the choice $\alpha' = \beta'/u'$ where $\beta'\in\ZZ$ and $u'$ is a divisor of $n$, 
$\ell=\sgn(m)(1-\alpha' n)$ is always possible. 
When $h=|m| \pm 1$, let $g=\gcd(m,n)$. If $(m,n/g)$ are coprimes, there is a solution given by $\alpha'=\beta'/g$ and  $\ell=\beta h$ where $(\beta,\beta')$ are B\'ezout coefficients of $(m,n/g)$. In both cases, one gets \eqref{eq:abel1} with $\lambda=1-n\alpha'$.
\end{itemize}
We now come to the case $|m|=|n|$. When $m=n$, cross-cancellations do not seem to occur, hence only the previous cases are to be considered. The case $m+n=0$ requires special attention for two reasons. 

First, the surface condition reads $c=\dfrac{N}{n}=-\dfrac{N}{m}$ and does not involve $p$. Matching 
as before the numerators and the denominators in \eqref{eq:Ymn-alt} leads to $-p^{\frac{1}{2}}=q^{N\alpha}$ where 
$\alpha \in \ZZ/2$ or $\alpha \in \ZZ/u$, $u$ is a divisor of $k_{\sup}$ or $k_{\sup}+1$; $-p^{*\frac{1}{2}}=q^{N\alpha'}$ 
where $\alpha' \in \ZZ/2$ or $\alpha' \in \ZZ/u'$, $u'$ is a divisor of $k'_{\sup}$ or $k'_{\sup}+1$. Given the value of $c$, 
this implies $\alpha-\alpha'=1/n$, which requires in fact $\alpha=\beta/n$, $\alpha'=(\beta-1)/n$. 
One recovers \eqref{eq:abel1} with $\lambda=\beta$.

Second, "cross-cancellations" between $p$ and $p^*$ shifted terms may now occur in \eqref{eq:Ymn-alt}. They can be consistently implemented when $n$ is odd. Following the same lines as in the previous cases, they yield \eqref{eq:abel4}.  

\section{Conclusion\label{sect:conclu}}

We have established a general pattern of construction for quadratic deformations of the Virasoro algebra, stemming from elliptic algebra structures. We have 
applied this pattern to the canonical $\gelpn$ algebra and to the alternative \ellipt{N} algebra defined in section \ref{sect:vertex}. This now opens several avenues of investigation, amongst which the most immediate will be to extend our scheme to higher-spin operators (i.e. operators involving higher powers of the $L$ matrices). This aims at getting deformations of the $W_N$ algebras, generalizing our past results in \cite{AFRS97b} on lines similar to the current work. 
The issue of further extending our approach to construct quadratic algebras, starting this time from generators of the dynamical elliptic algebras \cite{ABRR,JKOS} is 
technically trickier due to the
more intricate nature of the crossing/unitarity relations, but nevertheless quite promising.

Another subtle question is raised by our derivation of the original DVA \cite{SKAO} from the elliptic algebra \ellipt{N}. The exact meaning of this algebraic
structure is yet unclear. Note that the unitary elliptic R matrix is the one parametrizing the ZF algebra of type I 
vertex operators  introduced in \cite{FIJKMY95}.  Moreover it must be again emphasized that the original construction
in \cite{SKAO} relied on the use of difference operators acting naturally on a basis of Macdonald polynomials \cite{McD}. Macdonald polynomials may therefore be key objects
in understanding the nature of the \ellipt{N} algebra. Conversely this also suggests that an alternative construction of our DVA obtained from $\gelpn$, this time mimicking \cite{SKAO}
by using difference operators, may be available. This would then naturally raise a further question: which polynomial functions would  be relevant, again Macdonald polynomials or some other special functions?

\subsection*{Acknowledgments}
J.A. wishes to thank LAPTh Annecy for their kind hospitality. 

\appendix

\section{Jacobi theta functions\label{app:A}}

Let $\HH = \{ z\in\CC \,\vert\, \mbox{Im} z > 0 \}$ be the upper half-plane and 
$\Lambda_\tau = \{ \lambda_1\,\tau + \lambda_2 \,\vert\, \lambda_1,\lambda_2 \in \ZZ \,, \tau \in \HH \}$ 
the lattice with basis $(1,\tau)$ in the complex plane. 
One denotes the congruence ring modulo $N$ by $\ZZ_N \equiv \ZZ/N\ZZ$ with basis $\{0,1,\dots,N-1\}$. 
One sets $\omega = e^{2i\pi/N}$.  
Finally, for any pairs $\gamma=(\gamma_1,\gamma_2)$ and $\lambda=(\lambda_1,\lambda_2)$ of numbers, we define the
(skew-symmetric) pairing $\langle\gamma,\lambda\rangle \equiv \gamma_1\lambda_2 - \gamma_2\lambda_1$.

One defines the Jacobi theta functions with rational characteristics
$\gamma=(\gamma_1,\gamma_2) \in \sfrac{1}{N} \ZZ \times \sfrac{1}{N} \ZZ$ by:
\begin{equation}
\vartheta\car{\gamma_1}{\gamma_2}(\xi,\tau) = \sum_{m \in \ZZ} 
\exp\Big(i\pi(m+\gamma_1)^2\tau + 2i\pi(m+\gamma_1)(\xi+\gamma_2) \Big) \,.
\end{equation}
The functions $\vartheta\car{\gamma_1}{\gamma_2}(\xi,\tau)$ satisfy the following shift properties:
\begin{align}
& \vartheta\car{\gamma_1+\lambda_1}{\gamma_2+\lambda_2}(\xi,\tau) =
\exp(2i\pi \gamma_1\lambda_2) \,\, \vartheta\car{\gamma_1}{\gamma_2}(\xi,\tau) \,, \\
& \vartheta\car{\gamma_1}{\gamma_2}(\xi+\lambda_1\tau+\lambda_2,\tau) =
\exp(-i\pi\lambda_1^2\tau-2i\pi\lambda_1\xi) \, \exp(2i\pi\langle\gamma,\lambda\rangle) \,
\vartheta\car{\gamma_1}{\gamma_2}(\xi,\tau) \,,
\end{align}
where $\gamma=(\gamma_1,\gamma_2) \in \sfrac{1}{N}\ZZ \times
\sfrac{1}{N}\ZZ$ and $\lambda=(\lambda_1,\lambda_2) \in \ZZ \times \ZZ$. \\
Moreover, for arbitrary $\lambda=(\lambda_1,\lambda_2)$ (not necessarily integers), one has the following shift exchange:
\begin{equation}
\vartheta\car{\gamma_1}{\gamma_2}(\xi+\lambda_1\tau+\lambda_2,\tau) =
\exp(-i\pi\lambda_1^2\tau-2i\pi\lambda_1(\xi+\gamma_2+\lambda_2)) \,
\vartheta\car{\gamma_1+\lambda_1}{\gamma_2+\lambda_2}(\xi,\tau) \,.
\end{equation}

Considering the Jacobi $\Theta$ function:
\begin{equation}\label{theta:prod}
\Theta_p(z) = (z;p)_\infty \, (pz^{-1};p)_\infty \, (p;p)_\infty \,,
\end{equation}
where the infinite multiple products are defined by:
\begin{equation}
(z;p_1,\dots,p_m)_\infty = \prod_{n_i \ge 0} (1-z p_1^{n_1} \dots p_m^{n_m}) \,,
\end{equation}
the Jacobi theta functions with rational characteristics
$(\gamma_1,\gamma_2) \in \sfrac{1}{N} \ZZ \times \sfrac{1}{N} \ZZ$ can be expressed in terms of the Jacobi $\Theta$ functions as:
\begin{equation}
\vartheta\car{\gamma_1}{\gamma_2}(\xi,\tau) = (-1)^{2\gamma_1\gamma_2} \, p^{\frac{1}{2}\gamma_1^2} \, z^{2\gamma_1} \,
\Theta_{p}(-e^{2i\pi\gamma_2} p^{\gamma_1+\frac{1}{2}} z^2) \,,
\end{equation}
where $p = e^{2i\pi\tau}$ and $z = e^{i\pi \xi}$. \\
It is easy to show that the $\Theta_{a^2}(z)$ function enjoys the following properties:
\begin{equation}\label{id:theta}
\Theta_{a^2}(a^2z) = \Theta_{a^2}(z^{-1}) = -\frac{\Theta_{a^2}(z)}{z}
\mb{and} \Theta_{a^2}(az) = \Theta_{a^2}(az^{-1}).
\end{equation}

\section{Definition of the $N$-elliptic $R$-matrix\label{app:B}}

The starting point of the definition of the elliptic quantum algebras of vertex type is 
the $N$-elliptic $R$-matrix in $\mbox{End}(\CC^N) \otimes \mbox{End}(\CC^N)$ associated 
to the $\ZZ_{N}$-vertex model \cite{Bela,ChCh}, given by
\begin{equation}
\mathcal{Z}(z,q,p) = z^{2/N-2} \frac{1}{\kappa(z^2)} \frac{\vartheta\car{\half}{\half}(\zeta,\tau)} 
{\vartheta\car{\half}{\half}(\xi+\zeta,\tau)} \,\, 
\sum_{(\alpha_1,\alpha_2)\in\ZZ_N\times\ZZ_N} 
W_{(\alpha_1,\alpha_2)}(\xi,\zeta,\tau) \,\, I_{(\alpha_1,\alpha_2)} \otimes I_{(\alpha_1,\alpha_2)}^{-1} \,,
\end{equation}
where the variables $z,q,p$ are related to the variables $\xi,\zeta,\tau$ by
\begin{equation}
z=e^{i\pi\xi} \,,\qquad q=e^{i\pi\zeta} \,,\qquad p=e^{2i\pi\tau} \,.
\end{equation}
We set
\begin{equation}
R(z,q,p) = (g^{\frac{1}{2}} \otimes g^{\frac{1}{2}}) \mathcal{Z}(z,q,p) (g^{-\frac{1}{2}} \otimes g^{-\frac{1}{2}}) \,,
\label{eq220}
\end{equation}
where the $N \times N$ matrix $g$ is defined by
\beq\label{def:gij}
g_{ij} = \omega^i\delta_{ij}\,,\qquad 1\leq i,j\leq N \mb{with} \omega = e^{2i\pi/N}.
\eeq
The normalization factor is chosen as follows:
\begin{equation}
\label{eq:kappa}
\frac{1}{\kappa(z^2)} = \frac{(q^{2N}z^{-2};p,q^{2N})_\infty \, (q^2z^2;p,q^{2N})_\infty \, (pz^{-2};p,q^{2N})_\infty \,
(pq^{2N-2}z^2;p,q^{2N})_\infty} {(q^{2N}z^2;p,q^{2N})_\infty \, (q^2z^{-2};p,q^{2N})_\infty \, (pz^2;p,q^{2N})_\infty \,
(pq^{2N-2}z^{-2};p,q^{2N})_\infty} \,.
\end{equation}
The Jacobi theta functions and the infinite products are defined in Appendix \ref{app:A}. \\
The functions $W_{(\alpha_1,\alpha_2)}$ are given by
\begin{equation}
W_{(\alpha_1,\alpha_2)}(\xi,\zeta,\tau) = \frac{\vartheta\car{\sfrac{1}{2}+\alpha_1/N}
{\sfrac{1}{2}+\alpha_2/N}(\xi+\zeta/N,\tau)}{N\vartheta\car{\sfrac{1}{2}+\alpha_1/N}
{\sfrac{1}{2}+\alpha_2/N}(\zeta/N,\tau)} 
\end{equation}
and $I_{(\alpha_1,\alpha_2)} = g^{\alpha_2} \, h^{\alpha_1}$ where the $N \times N$ matrix $h$ is such that 
\beq\label{def:h}
h_{ij} = \delta_{i+1,j}\,, \qquad 1\leq i,j\leq N,
\eeq
the addition of indices being understood modulo $N$.

We recall the following proposition \cite{AFRS99}, see also \cite{Tra85,RT86}:
\begin{prop}\label{propone}
The matrix $R(z) \equiv R(z,q,p)$ satisfies the following properties: \\
-- Yang--Baxter equation:
\begin{equation}
R_{12}(z) \, R_{13}(w) \, R_{23}(w/z) = R_{23}(w/z) \, R_{13}(w) \, R_{12}(z) \,,
\label{eq221}
\end{equation}
-- Unitarity:
\begin{equation}
R_{12}(z) \, R_{21}(z^{-1}) = 1 \,,
\label{eq222}
\end{equation}
-- Regularity ($P_{12}$ is the permutation matrix):
\begin{equation}
R_{12}(1) = P_{12} \,,
\end{equation}
-- Crossing-symmetry:
\begin{equation}
R_{12}(z)^{t_2} \, R_{21}(z^{-1}q^{-N})^{t_2} = 1 \,,
\label{eq223}
\end{equation}
-- Antisymmetry:
\begin{equation}
R_{12}(-z) = \omega \, (g^{-1} \otimes \II) \, R_{12}(z) \, (g \otimes \II) \,,
\label{eq224}
\end{equation}
-- Quasi-periodicity:
\begin{equation}
\widehat R_{12}(-z p^{\frac{1}{2}}) = (g^{\frac{1}{2}} h g^{\frac{1}{2}} \otimes \II)^{-1} \, \widehat R_{21}(z^{-1})^{-1} \, (g^{\frac{1}{2}} h g^{\frac{1}{2}} \otimes \II) \,,
\label{eq225}
\end{equation}
where
\begin{equation}
\widehat R_{12}(z) \equiv \widehat R_{12}(z,q,p) = \tau_N(q^{\frac 12}z^{-1}) \, R_{12}(z,q,p) \,,
\label{eq226}
\end{equation}
the function $\tau_N(z)$ being defined by
\begin{equation}
\tau_N(z) = z^{\frac{2}{N}-2} \, \frac{\Theta_{q^{2N}}(qz^2)}{\Theta_{q^{2N}}(qz^{-2})} \,.
\label{eq227}
\end{equation}
The function $\tau_N(z)$ is $q^N$-periodic, $\tau_N(q^Nz) = \tau_N(z)$, and satisfies $\tau_N(z^{-1}) = \tau_N(z)^{-1}$.
\end{prop}

{\bf Remark}: The crossing-symmetry and the unitarity properties of $R_{12}$ allow to exchange the inversion and the
transposition when applied to the matrix $R_{12}$ as (the same property also holds for the matrix $\widehat R_{12}$):
\begin{equation}
\Big(R_{12}(x)^{t_2}\Big)^{-1} = \Big(R_{12}(q^Nx)^{-1}\Big)^{t_2} \,.
\label{eq230}
\end{equation}
Note also that the unitarity property for $\widehat R_{12}$ now reads 
\begin{equation}
\widehat R_{12}(z) \, \widehat R_{21}(z^{-1}) = \tau_N(q^{\frac 12}z) \, \tau_N(q^{\frac 12}z^{-1}) \equiv \mathcal{U}(z).
\label{eq:unitarity}
\end{equation}
The function $\cU(z)$ is extensively used in our discussion. In term of Jacobi $\Theta$ functions, it reads
\beq\label{def:U}
\cU(z)=q^{\frac2N-2}\,\frac{\Theta_{q^{2N}}(q^2z^2) \, \Theta_{q^{2N}}(q^2z^{-2})} {\Theta_{q^{2N}}(z^2)\Theta_{q^{2N}}(z^{-2})} .
\eeq
\paragraph{In the case $N=2$.} The elliptic $R$-matrix takes the following form:
\begin{equation}
\label{eq:aqpgl2}
R_{12}(z,q,p) = \frac{1}{\kappa(z^2)} \frac{(p^2;p^2)_\infty} {(p;p)_\infty^2} \left(
\begin{array}{cccc}
a(z) & 0 & 0 & d(z) \\
0 & b(z) & c(z) & 0 \\
0 & c(z) & b(z) & 0 \\
d(z) & 0 & 0 & a(z) \\
\end{array}
\right)\,,
\end{equation}
where
\begin{equation}
\begin{split}
& a(z) = z^{-1} \; \frac{\Theta_{p^2}(pz^2) \; \Theta_{p^2}(pq^2)} {\Theta_{p^2}(pq^2z^2)} \,,
\qquad d(z) = -\frac{p^{1/2}}{q z^{2}} \; \frac{\Theta_{p^2}(z^2) \; \Theta_{p^2}(q^2)} {\Theta_{p^2}(pq^2z^2)}\,, \\ 
& b(z) = qz^{-1} \; \frac{\Theta_{p^2}(z^2) \; \Theta_{p^2}(pq^2)} {\Theta_{p^2}(q^2z^2)} \,,
\qquad c(z) = \frac{\Theta_{p^2}(pz^2)\Theta_{p^2}(q^2)} {\Theta_{p^2}(q^2z^2)}\,.
\end{split}
\end{equation}

\section{The quantum determinant in $\gelp$\label{sect:qdet}}

The elliptic $R$-matrix \eqref{eq:aqpgl2} satisfies $R_{12}(-q^{-1}) = 1-P_{12}$ ($P_{12}$ being the permutation matrix), which is equal to the antisymmetrizer $A_2$ in $(\CC^2)^{\otimes 2}$:
\begin{equation}
A_2 (e_{i_1} \otimes e_{i_2}) = \sum_{\sigma\in\mathfrak{S}_2} (\sgn\sigma) \; e_{i_{\sigma(1)}} \otimes e_{i_{\sigma(2)}}
\end{equation}
where $(e_{1},e_{2})$ denotes the canonical basis of $\CC^2$.

Since the antisymmetrizer $A_2$ is a one-dimensional operator in $(\CC^2)^{\otimes 2}$, the $RLL$ relations allows one to define the quantum determinant $\qdet L(z)$ \cite{Molev} such that
\begin{equation}
R_{12}(q^{-1}) L_1(z) L_2(q^{-1}z) = L_2(q^{-1}z) L_1(z) R_{12}(q^{-1}) = \qdet L(z) R_{12}(q^{-1})
\end{equation}
Explicitly, one gets
\begin{equation}
\begin{split}
\qdet L(z) &= L_{11}(q^{-1}z) L_{22}(z) - L_{21}(q^{-1}z) L_{12}(z) \\
&= L_{22}(q^{-1}z) L_{11}(z) - L_{12}(q^{-1}z) L_{21}(z) \\
&= L_{11}(z) L_{22}(q^{-1}z) - L_{12}(z) L_{21}(q^{-1}z) \\
&= L_{22}(z) L_{11}(q^{-1}z) - L_{21}(z) L_{12}(q^{-1}z) 
\end{split}
\end{equation}
It can be shown that the quantum determinant $\qdet L(z)$ lies in the center of $\gelp$ \cite{FIJKMY95}. \\
The elliptic quantum algebra $\elp$ \cite{FIJKMY,JKOS} is then defined as the quotient algebra
\begin{equation}
\elp = \gelp/\langle \qdet L(z) - q^{\frac{c}{2}} \rangle \,.
\end{equation}
The quantum comatrix $\widehat{L}(z)$ is defined in $\gelp$ by
\begin{equation}
\widehat{L}(q^{-1}z) L(z) = \qdet L(z)\,\II_2 \,.
\label{eq:comat}
\end{equation}
Explicitly, one gets
\begin{equation}
\widehat{L}(z) = \begin{pmatrix} \phantom{-}L_{22}(z) & -L_{12}(z) \\ -L_{21}(z) & \phantom{-}L_{11}(z) \end{pmatrix}
\label{eq:lchapeau}
\end{equation}
It follows from \eqref{eq:comat} that $L^{-1}(z) = \big(\qdet L(z)\big)^{-1} \; \widehat{L}(q^{-1}z)$.

\end{document}